\documentclass[11pt,twoside]{article}
\usepackage{cozumel2005}
\usepackage{epsf}
\usepackage{psfig}
\usepackage{graphicx}
\usepackage{lscape}
\begin{document}
\title{Deriving star formation histories from integrated light: Colors and indices}    
\author{Thomas Lilly and Uta Fritze-v. Alvensleben}   
\affil{Institut f\"ur Astrophysik, Universit\"at G\"ottingen, Germany}    

\begin{abstract} 
We present results of a detailed study aiming at understanding to what
precision star formation histories (SFHs) can be determined for distant
galaxies observable in integrated light only. Using our evolutionary
synthesis code, we have performed a set of simulations of galaxies with a
wide range of different SFHs. By analysing the resulting colors, spectra
and Lick indices, we investigate to which extent different SF scenarios can
be discriminated on the basis of their photometric and spectral properties,
respectively.
We find the robust result that no later than 4 Gyrs after the latest
episode of enhanced star formation all scenarios exhibit very similar
colors and indices; in practice, it is not possible to distinguish
different scenarios of star formation which have evolved for more than 1,
at the utmost 3-4 Gyrs since the last star forming event, even when using
spectral indices. For how long different SF scenarios can be disentangled
highly depends on the range of colors available and absorption lines
considered, as well as on the details of the SFHs to be compared.
\end{abstract}




\section{Introduction}

Methods aiming at the reconstruction of the star formation history (SFH) of a galaxy are important tools for
understanding the evolution of these objects.
The available methods can be devided into two groups: Methods using color magnitude diagrams (CMDs) are widely
regarded as the most reliable ones. However, they require that the stellar population can be resolved into
individual stars; therefore, methods of this kind are limited to nearby galaxies, and there only to star fields
without crowding.
The second group of methods makes use of integrated-light properties of unresolved galaxies like colors and
spectra, which are far easier to get for a much wider range of galaxies, even for those at high redshift. 
Integrated colors and spectra can be analysed and interpreted by means of either population synthesis codes or
evolutionary synthesis codes like the GALEV code used in this paper.

Much work has been done in testing these codes with respect to internal errors, to the influence of the input
physics used, and to the degree of agreement/disagreement between different models (see, e.g.: Charlot et al.
1996; Cervi\~no et al. 2000 et seqq.; Bruzual 2001; Yi 2002). In addition, differences between observed and
model-predicted properties of galaxies and star clusters have been investigated in detail (see, e.g.: Vazdekis et
al. 2001; Schulz et al. 2002).\\

In this contribution, we deal with a quite different but related question: While ignoring the already well
explored model-dependent uncertainties mentioned above, we want to answer the fundamental question, to what
precision SFHs of galaxies can \emph{in principle} be determined by using integrated light only, that is,
independent of the model but dependent on the physical properties of a galaxy's stellar population.

Using our evolutionary synthesis code GALEV, we have therefore performed a set of simulations of galaxies
with a wide range of different SFHs but well defined and uniform input physics. That way, we restrict our
investigation to influences of the variation of the SFH on the spectro-photometric properties of a galaxy;
other parameters like the initial mass function or the metallicity are kept constant in this study.

By confrontation of the evolution of the colors and spectra resulting from the various simulations we then
investigate, to what extent different scenarios of star formation can be discriminated at all and as a function
of lookback time, and how this depends on the specific shapes of the assumed SFHs.
For clarity we restrict our study to simplified SFHs with constant star formation rates (SFRs) over various periods
of time, from long intervals of low SFR to starburst periods of high SFR.
Because the models are all synthesised using one genuine code and one genuine set of input physics, we
can perform this comparative study in a self-consistent way.\\

It should be kept in mind that, due to the character of the study, it is by no means to be regarded as a
practical instruction of how to recover the SFH of a given galaxy, although, as we hope, much can be learned
about how to do this most efficiently, e.g. by adressing the question what filters or spectral lines should be
analysed, and in how far the results can be reliably interpreted.\\
%

\section{A very short introduction to GALEV}

The evolutionary synthesis model GALEV simulates the spectrophotometric evolution of the integrated light of large
stellar populations like galaxies or star clusters. They are treated using a 1-zone model, i.e. their dynamical
properties remain unconsidered. As for all evolutionary synthesis models, the (historical) basis of the code are
the equations given by Tinsley (e.g., 1980), which describe the global balance of stars and gas in a galaxy, and
kind of a book-keeping algorithm that keeps track of all stars in a model galaxy at various timesteps and their
distribution over the HR diagram.

Input physics for the code include the theoretical spectral library from Lejeune et al. (1997, 1998) as well as
theoretical isochrones from the Padova group like the ones described by Bertelli et al. (1994) for 5 different
metallicities $Z$=0.0004, 0.004, 0.008, 0.02 and 0.05, but in the version from November 1999 that includes the
TP-AGB phase of stellar evolution (as described in Schulz et al. 2002).
We assume a standard Salpeter (1955) initial mass function (IMF) from 0.15 to about 70 M$_\odot$; the lowest mass
stars (M$_\odot < 0.6$) are taken from Chabrier \& Baraffe (1997) (cf. Schulz et al. 2002 for details).
As the basis for our models for Lick indices we employ the polynomial fitting functions of Worthey et al. (1994)
and Worthey \& Ottaviani (1997), which give Lick index strenghts of individual stars as a function of their
effective temperature $T_{\rm eff}$, surface gravity $g$, and metallicity [Fe/H]. Worthey et al. have calibrated
their fitting functions empirically using Milky Way stars.

Once an IMF is assumed, the basic free parameters of our models are the star formation rate (SFR) and the
metallicity of the stellar population.
The code then produces the time evolution (4Myr ... 16Gyr) of spectra (90\AA\ ... 160$\mu$m), colors in many filter
systems, and 25 Lick/IDS spectral indices.

For an exhaustive description of the code and its input physics see Schulz et al. (2002), Bicker et al. (2003),
Anders \& Fritze -- v. Alvensleben (2003), and Lilly \& Fritze -- v. Alvensleben (2005b).\\
%

\section{Scenarios in integrated light}

We want to explore in how far different SF scenarios can in retrospect be discriminated against each other.
Therefore, we have performed a set of simulations of galaxies with a range of - not realistic, but instructive -
SFHs. Other parameters are kept constant; for all scenarios, we assume a Salpeter(1955) IMF and a fixed metallicity
of $Z$ = 0.008.

For simplicity and lack of better knowledge about realistic burst shapes we assume rectangular burst shapes on top
of constant SF (we do not expect this simplification to substantially affect on results). Unless stated otherwise,
we assume that \emph{any two SF scenarios we compare have produced the same amount of stars in total}. Therefore,
the absolute numerical value of the SFR does not matter for the colors or spectral indices of the integrated light.
It is the \emph{relative distribution} of the SFR over the evolutionary time of the galaxy that is significant.

\begin{table}
   \caption{Characteristics of the SF scenarios to be compared; the enhancement of SF always refers to a
   ``basic rate'' of 1 M$_\odot$/yr.}
   \smallskip
\begin{center}
{\small
\begin{tabular}{c p{10.5cm}}
         \tableline
         \noalign{\smallskip}
         scenario & description\\
         \noalign{\smallskip}
         \tableline
         \noalign{\smallskip}
         0 & slightly enhanced SF (factor 3.4) from $t = 0$ Gyr through $t = 6.4$ Gyr\\
         1 & 2 bursts (SF enhanced by a factor of 20) of duration 0.4 Gyr at $t = 4$ Gyr and $t = 6$ Gyr\\
         2 & enhanced SF (factor 7) from $t = 4$ Gyr through $t = 6.4$ Gyr\\
         3 & 2 bursts (SF enhanced by a factor of 20) of duration 0.4 Gyr at $t = 0$ Gyr and $t = 6$ Gyr\\
         4 & 1 bursts (SF enhanced by a factor of 20) of duration 0.8 Gyr at $t = 5.6$ Gyr\\
         6 & 1 strong burst (SF enhanced by a factor of 40) of duration 0.4 Gyr at $t = 6$ Gyr\\
         \noalign{\smallskip}
         \tableline
\end{tabular}
}
\end{center}
\label{tab.sfhs}
\end{table}

The characteristics of all SF scenarios compared in this paper can be found in Table \ref{tab.sfhs}.
Note that, throughout this paper, age follows galactic evolution (not lookback time); thus, a galaxy age of 6 Gyr
refers to a galaxy which has evolved for 6 Gyrs, starting its evolution at 0 Gyr.

Throughout this study, we assume a typical photometric accuracy of about 0.1 mag, at best 0.05 mag. This means, if
differences in model-predicted colors between different scenarios of SF do not exceed 0.1 mag, they will be
considered as observationally indistinguishable. In case of indices, this limit is assumed to be about 0.1\AA.\\
%

\subsection{Colors}

\begin{figure}[t]
   \plottwo{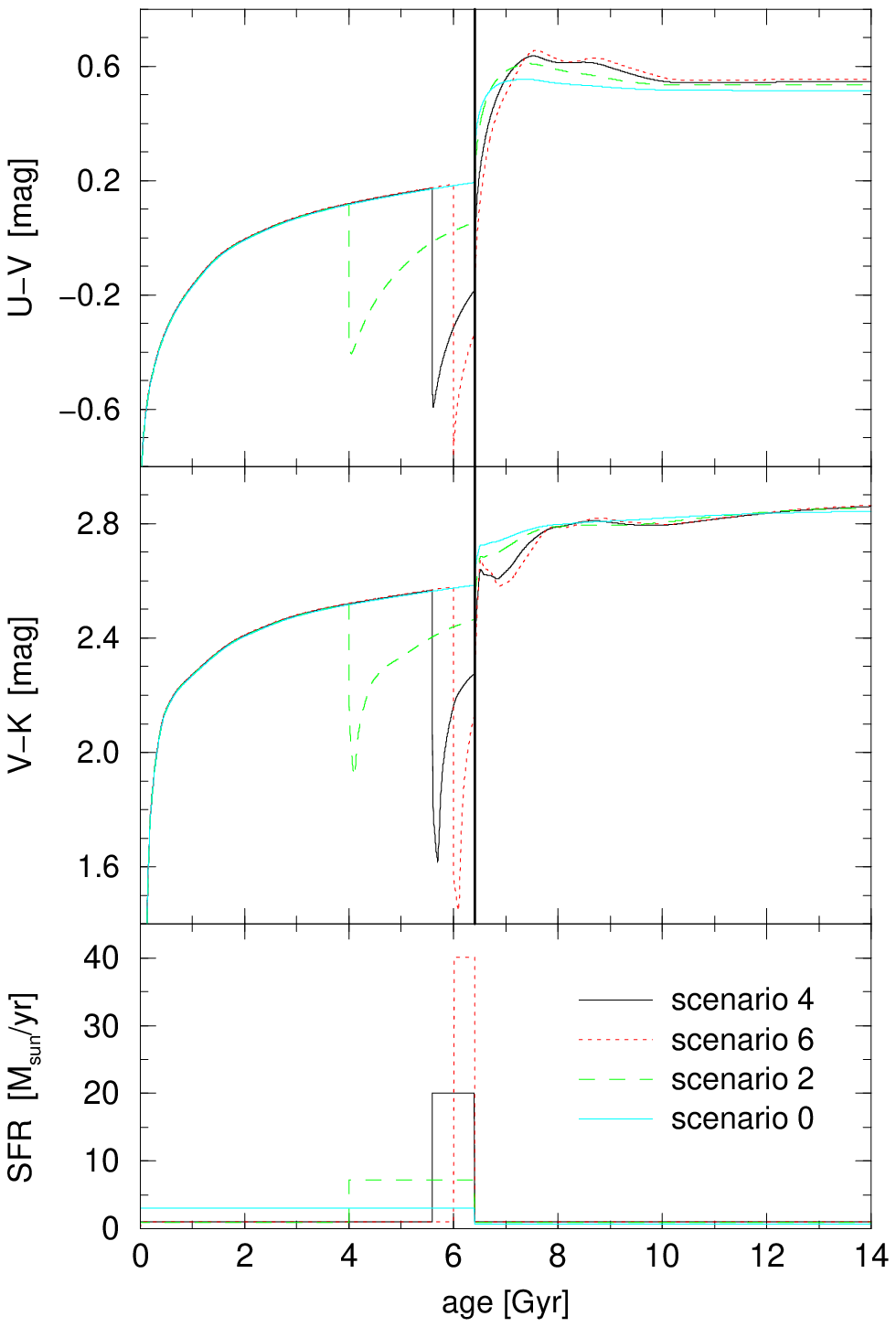}{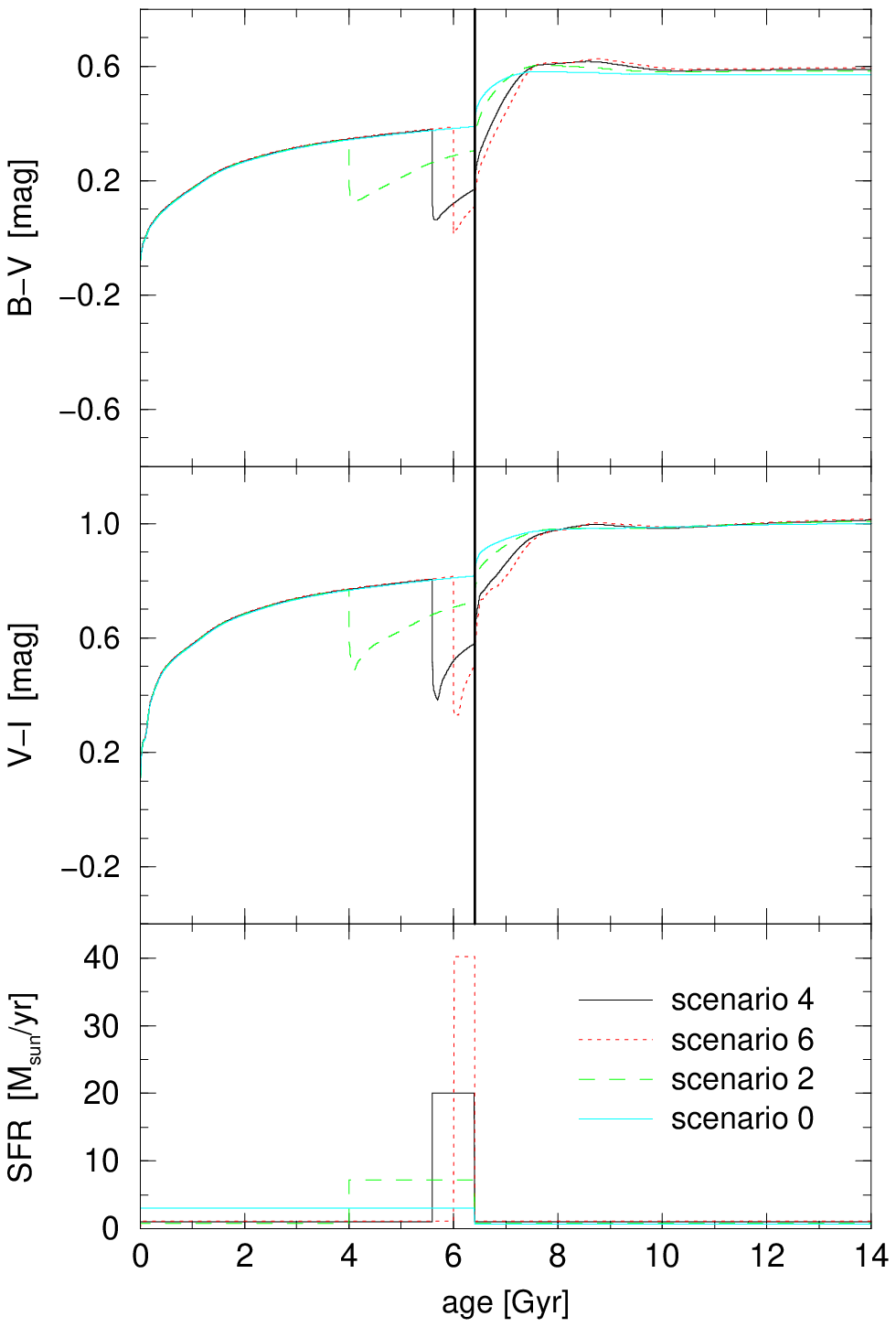}
   \caption{Confrontation of the photometric evolution of scenarios 0, 2, 4, 6 in terms of the colors U-V and V-K
   ({\itshape left\/}), and B-V and V-I ({\itshape right\/}). At 6.4 Gyr, when the latest phase of enhanced SF
   ends, a thick line is plotted. The bottom plots show the respective SFHs to be compared.}
   \label{abb.colors1}
\end{figure}

First, we confront the photometric evolution of four scenarios, each of them featuring 1 burst, ranging from a
strong burst at 6 Gyr galaxy age (scenario 6) through bursts with weaker strengths but longer durations (scenarios
4 and 2), to a phase of only slightly enhanced SF (scenario 0).
In each case, the bursts are put on top of a constant ``basic rate'' of SF of 1 M$_\odot$/yr, and all ``bursts''
finish at 6.4 Gyr galaxy age.
Figure \ref{abb.colors1} shows the photometric evolution of these scenarios. To guide the eye, we plot a thick line
at 6.4 Gyr. The bottom plots show the respective SFHs to be compared.

The plots show impressively that already 1 Gyr after the latest enhanced SF period all scenarios show nearly
identical colors.
Even though scenario 0 differs in U-V from the other scenarios for at least 4 Gyrs due to the larger amount of
still existing red giant stars originating from the phases of enhanced or 'bursty' SF in the latter, this
feature can hardly be used for the reconstruction of SFHs because the maximum difference of 0.1 mag declines
rapidly to practically indistinguishable values of about 0.05 mag.

Hence, the ``lookback time'', during which burst of different strengths can be discriminated in integrated light
is only 1, at the utmost 4 Gyr; after that time, it is not even possible to discriminate between a strong burst
scenario of galaxy evolution and a very ``quiet'' evolution like that of scenario 0, if no other clues about a
possibly violent history of the galaxy are available but only global broad-band colors.\\

Since any two SF scenarios we compare have produced the same amount of stars in total, the amount of long-living
low-mass stars is roughly the same in all scenarios after the most recent burst finishes; therefore, the
photometric distinction between different SF scenarios is mainly determined by luminous high-mass stars and their
evolution. During a burst, for example, according to the IMF many more cool and red low-mass stars are formed than
blue high-mass stars. However, due to their extremely high luminosities, and despite their considerably smaller
number, the massive stars dominate the integrated light, resulting in an abrupt change of color in the model
galaxy as soon as the burst starts.

As the bluest massive stars die out after the end of the burst, colors get redder very fast. This ``reddening''
can already be observed during bursts: Whereas the number of very blue and very massive stars reaches some
equilibrium early in the burst, the redder low-mass stars accumulate during the burst (as well as during the whole
lifetime of the galaxy) due to their considerably longer lifetimes, and therefore begin to overbalance the bluest
high-mass stars.

These effects can be observed in all colors; they are stronger in colors covering a large spectral range like U-V
or V-K than in colors from two close passbands like B-V and V-I.

\begin{figure}
   \begin{center}
   \includegraphics[width=0.41\linewidth]{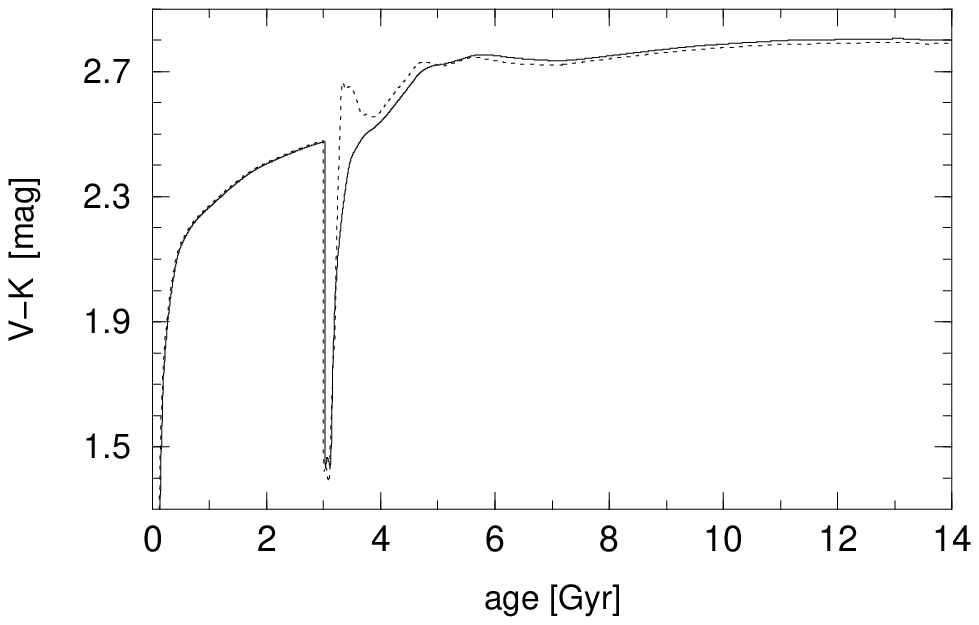}
   \includegraphics[width=0.41\linewidth]{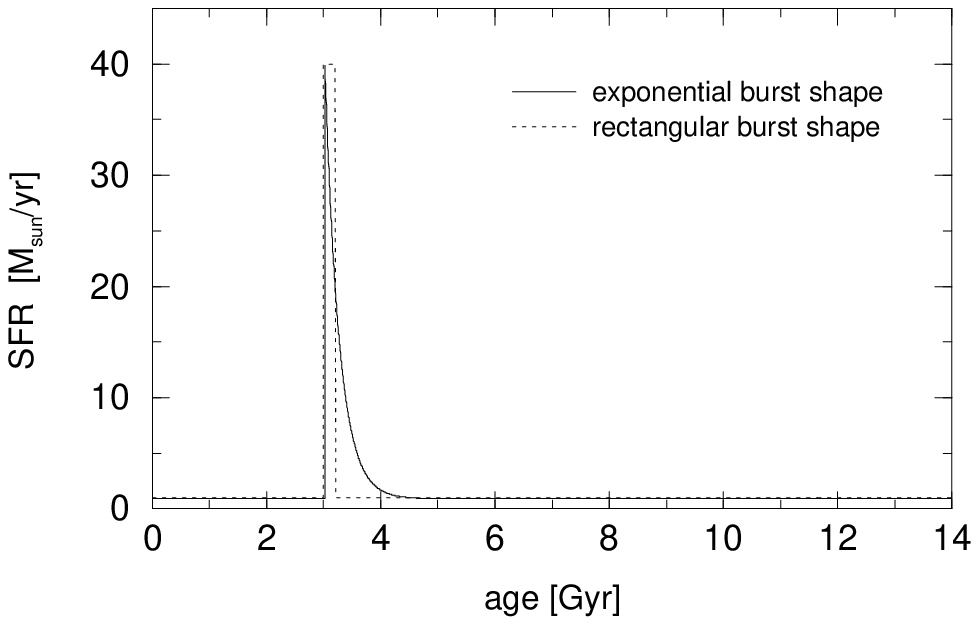}
   \caption{({\itshape Left:\/}) Photometric evolution in V-K of a 'realistic' starburst (solid line) as compared
   to a rectangular one (dotted line). ({\itshape Right:\/}) Respective SFHs. See text for details.}
   \label{abb.burst}
   \end{center}
\end{figure}

However, the difference between the scenarios is not larger in V-K than in V-I; this is explained by
intermediate-mass stars (stars with initial masses of $2 M_\odot \le m \le 7 M_\odot$) passing through the TP-AGB
(thermally pulsing asymthotic giant branch) phase from an age of $10^8$ yr up to an age of $10^9$ yr. During this
phase, they are located in the upper right of the HR diagram and account for approx. 40-60\% of the K-band light
(Lan{\c c}on 1999). After each burst a typical bump in V-K caused by TP-AGB stars can easily be identified; a
comparison with a model that does not include the TP-AGB phase can be found in Schulz et al. 2002 (Fig. 1).

Whether this bump occurs in V-K also depends on the shape of the declining phase of the burst. In Figure
\ref{abb.burst}, we compare an exponentially declining burst (decay time $\tau = 2.5 \cdot 10^8$ yr; for the burst
model implemented here cf. Bicker et al. 2002) with a rectangular shaped one of equal strength (lasting $\sim 210$
Myrs, so that the same number of stars is formed in both cases); both bursts start at a galaxy age of 3 Gyr.
Figure \ref{abb.burst} shows that in the case of an exponentially declining burst, the influence of the TP-AGB
phase is much weaker than in the 'abruptly declining' burst model used in the simulations for this paper,
resulting in a much better discernability between burst and non-burst scenarios during the declining phase.
This can be explained by considering that in the exponentially declining burst phase the photometric influence of
the TP-AGB stars is diluted by red supergiants still forming during the same period. Due to the extremely short
lifetimes of these very massive red supergiants, they are less important shortly after a rectangular burst.

Hence, in ``real'' galaxy evolution scenarios we expect V-K to be a better indicator of former SF as V-I. The
lookback time, however, over which any two SFHs can be discriminated from each other, is not longer in V-K than in
V-I.\\

\begin{figure}[t]
   \plottwo{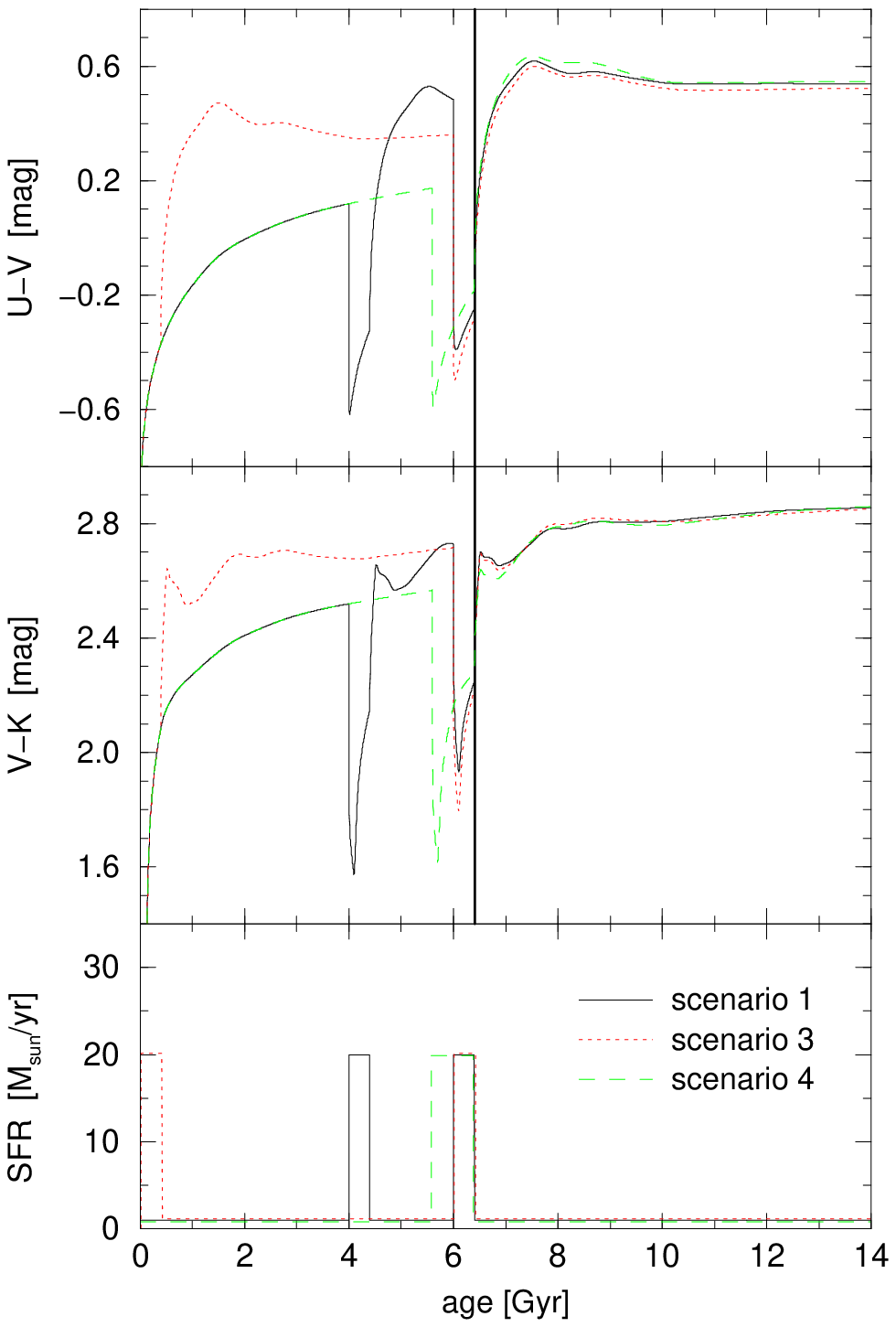}{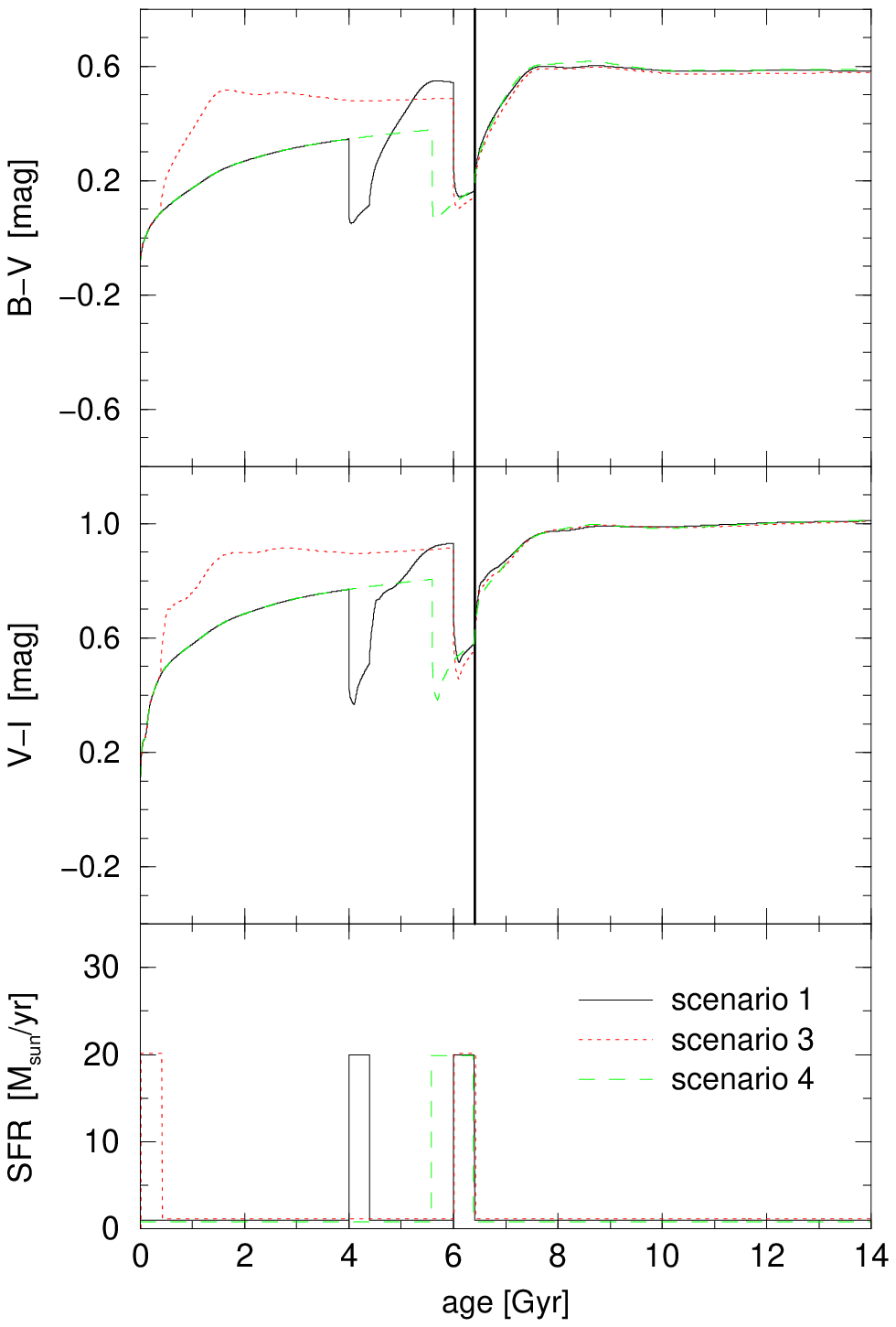}
   \caption{Confrontation of the photometric evolution of scenarios 1, 3, 4 in terms of the colors U-V and V-K
   ({\itshape left\/}), and B-V and V-I ({\itshape right\/}). At 6.4 Gyr, when the latest phase of enhanced SF
   ends, a thick line is plotted. The bottom plots show the respective SFHs to be compared.}
   \label{abb.colors2}
\end{figure}

In Figure \ref{abb.colors2} we confront the photometric evolution of three scenarios, each of them featuring a
burst going on between 6.0 and 6.4 Gyr with a SFR of 20 M$_\odot$/yr on top of galaxy models with different
former SF:
Scenarios 1 and 3 feature a previous burst with similar characteristics at a galaxy age of 4 Gyr, and at the
beginning of galactic evolution (0 Gyr galaxy age), respectively.
In scenario 4, the two bursts are replaced by one single burst lasting twice as long (i.e., for 800 Myrs),
starting at 5.6 Gyr galaxy age.

Using global broad-band colors alone, all these scenarios are practically indistinguishable directly after the end
of the most recent burst:
After a burst, we are no longer able to study the SFH prior this event.\\
%

\subsection{Indices}

\begin{figure}[!t]
   \plottwo{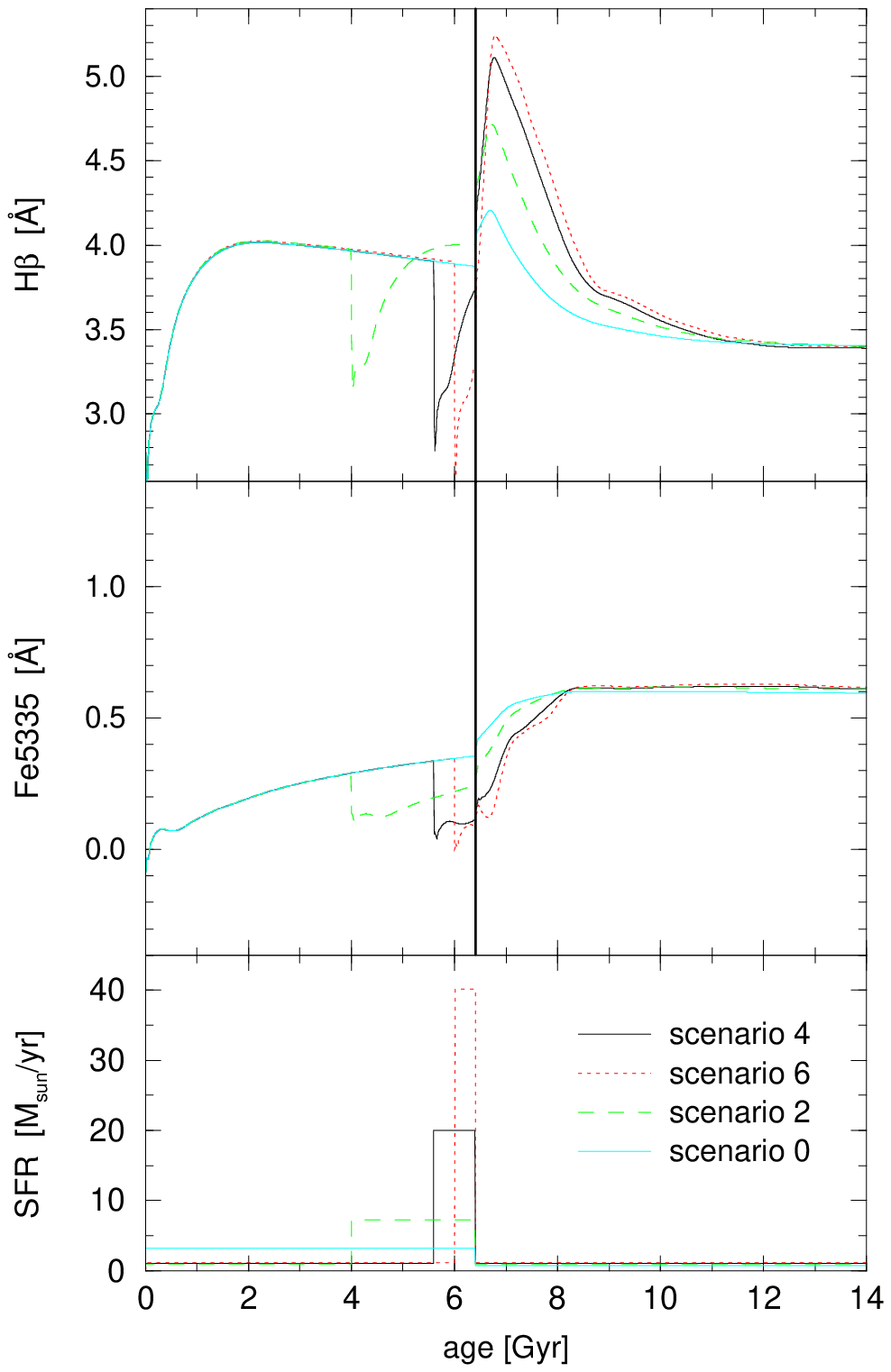}{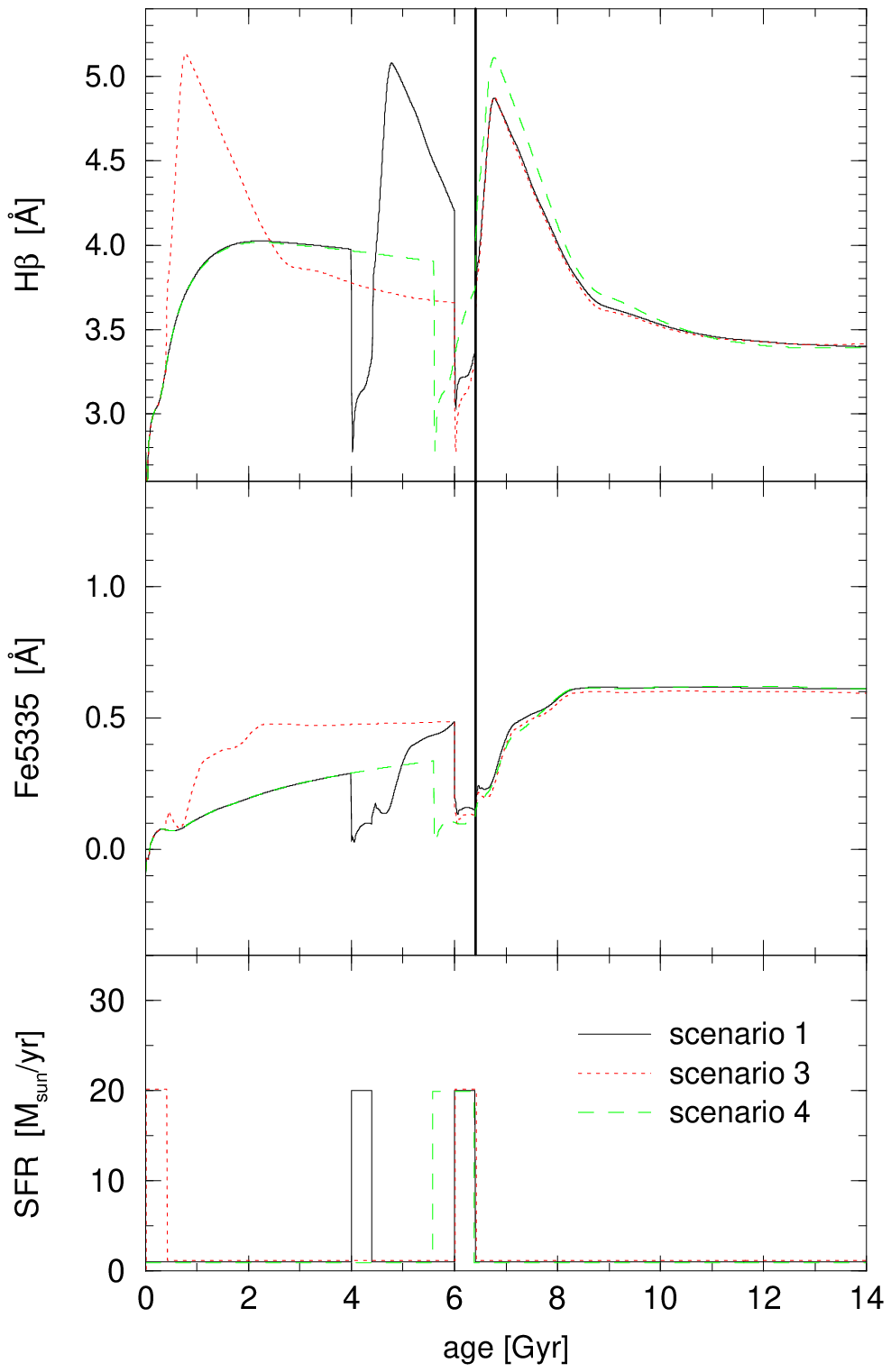}
   \caption{Confrontation of the evolution of scenarios 0, 2, 4, 6 ({\itshape left\/}) and 1, 3, 4 ({\itshape
   right\/}), respectively, in terms of the Lick indices H$\beta$ and Fe5335. At 6.4 Gyr, when the latest phase of
   enhanced SF ends, a thick line is plotted. The bottom plots show the respective SFHs to be compared.}
   \label{abb.2indices}
\end{figure}
\begin{figure}[!t]
   \begin{center}
   \includegraphics[width=0.65\linewidth]{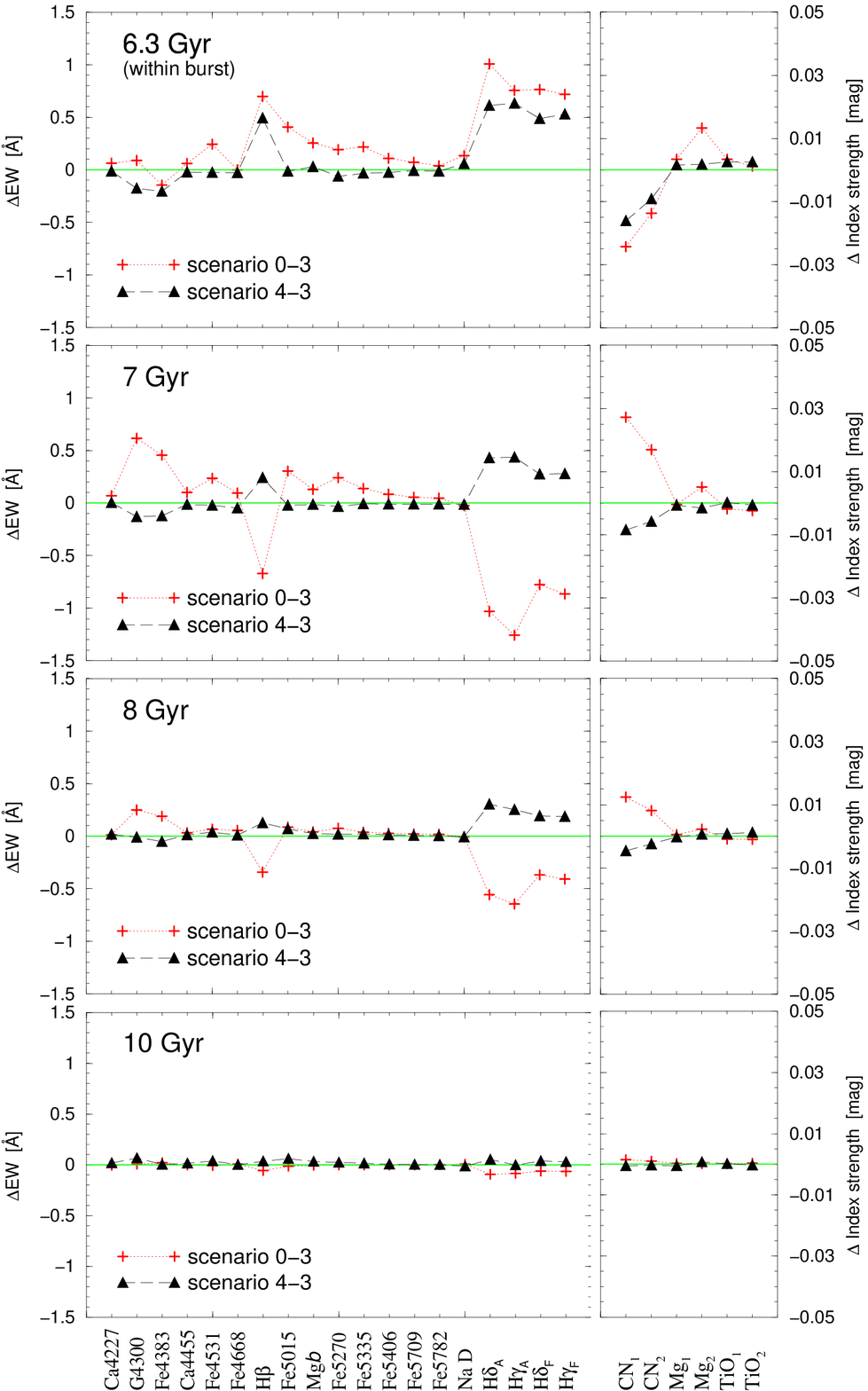}
   \includegraphics[width=0.45\linewidth]{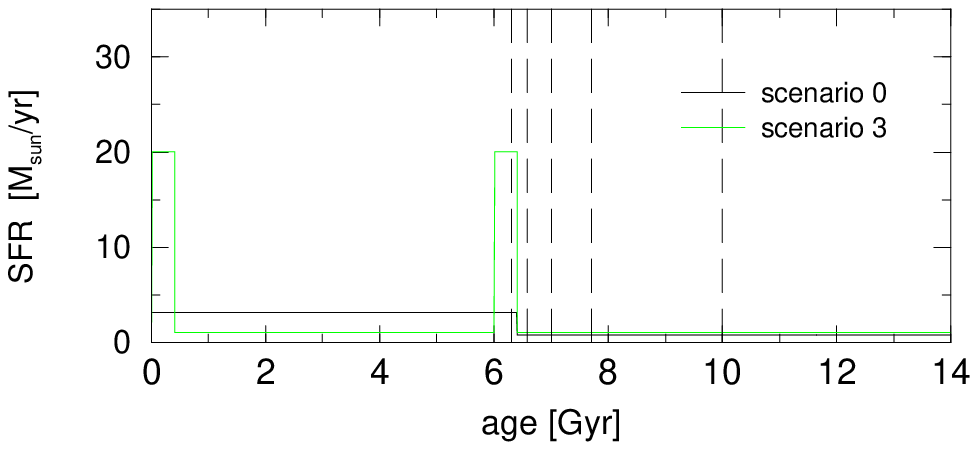}
   \includegraphics[width=0.45\linewidth]{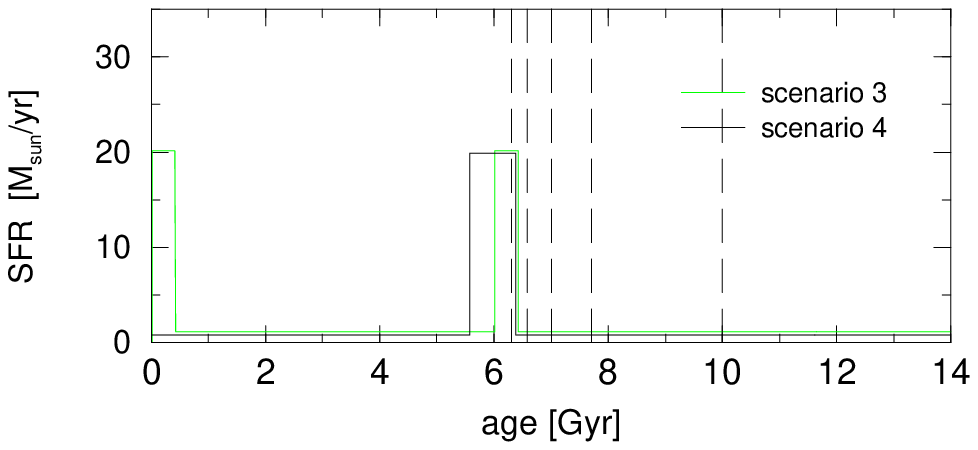}
   \caption{{\itshape Top:\/} Absolute differences of all 25 Lick indices (in terms of index strength) of the
   scenarios 0 and 3 and of the scenarios 4 and 3, respectively, for the four model ages indicated in the plot.
   {\itshape Bottom:\/} SFHs of the scenarios; the ages 6.3, 6.6, 7.0, 7.7 and 10 Gyr are indicated by
   dashed lines.}
   \label{abb.allindices}
   \end{center}
\end{figure}

In the last Section, we have seen that not even bursts with very different SFRs can be discriminated very well
for more than 1, at the utmost 4 Gyr of lookback time in terms of colors.
Surprisingly, the situation improves not very much if spectral indices are considered instead of broad-band
colors.\\

In Figure \ref{abb.2indices}, we again confront the evolution of scenarios featuring bursts of different strength
and duration (left panels; cf. last section, Fig. \ref{abb.colors1}), and of scenarios featuring bursts going on
between 6.0 and 6.4 Gyr with equal strenth each, but different former SF (right panels; cf. last section, Fig.
\ref{abb.colors2}), but this time in terms of the Lick index H$\beta$, which is known to be sensitive to age, and
of the Lick index Fe5335, which is known to be more sensitive to metallicity than to age (cf. Worthey 1994).
Since the scenarios vary only in SFH, but have fixed metallicity, the scenarios can be discriminated much
better in H$\beta$ than in the metallicity sensitive Fe5335, as expected.

As in case of colors, index strengths are very similar in scenarios 1, 3, 4 after the end of the last burst (Fig.
\ref{abb.2indices}, right panels). Only scenario 4, in which the most recent burst has twice the duration (800
Myrs) of those in the other scenarios, differs in H$\beta$ by about 0.2 \AA\ from the others during the first Gyr
after the end of the most recent burst. This difference in H$\beta$ is due to the lifetime of about $\frac{1}{2}$
to 1 Gyr of early A-type stars, which feature strong Balmer lines in their spectra.

The different bursts of scenarios 0, 2, 4, 6, however, can be discriminated much better using indices than using
colors (Fig. \ref{abb.2indices}, right panels); in our simulations, the difference between the strong burst
scenario 6 and the non-burst scenario 0 reaches more than 1 \AA\ in H$\beta$ during the first Gyr after the end of
the burst. One Gyr later, the difference has decreased to less than 0.5 \AA, but 4 Gyrs after the end of the burst
the difference is already too small to be measured in praxi.\\

We have compared the scenarios only in terms of H$\beta$ and Fe5335; in Figure \ref{abb.allindices}, we plot the
absolute difference in terms of index strength between different scenarios for \emph{all} Lick indices. For means
of clarity, we plot the values for four galaxy ages only (6.3 Gyr, wich is still during the most recent burst, 7
Gyr, 8 Gyr, and 10 Gyr), and for two pairs of scenarios: With scenarios 0 and 3 (crosses in Fig.
\ref{abb.allindices}) we confront a burst and a non-burst scenario, with scenarios 3 and 4 (triangles in Fig.
\ref{abb.allindices}) we confront a short burst scenario with a long burst scenario where both bursts form
identical amounts of stars in total.
(Note that we do not expect the first burst of scenario 3 to have any significant effect on index strengths at the
ages considered here; it only guarantees that all scenarios form equal amounts of stars in total.)

From these plots, the absolute sensitivity of individual Lick indices to age can directly be read off, showing for
example the large age-sensitivity of Balmer lines (however, note that the problem of age-metallicity degeneracy
is ignored by this approach).

At 10 Gyr galaxy age, i.e. less than 4 Gyrs after the most recent burst, the scenarios are practically
indistinguishable in all indices, with maximal differences of only about 0.1 \AA.
This shows that the results obtained earlier for H$\beta$ and Fe5335 are valid for all indices.\\
%

\subsection{Spectra}

\begin{figure}[!th]
   \begin{center}
   \includegraphics[width=1\linewidth]{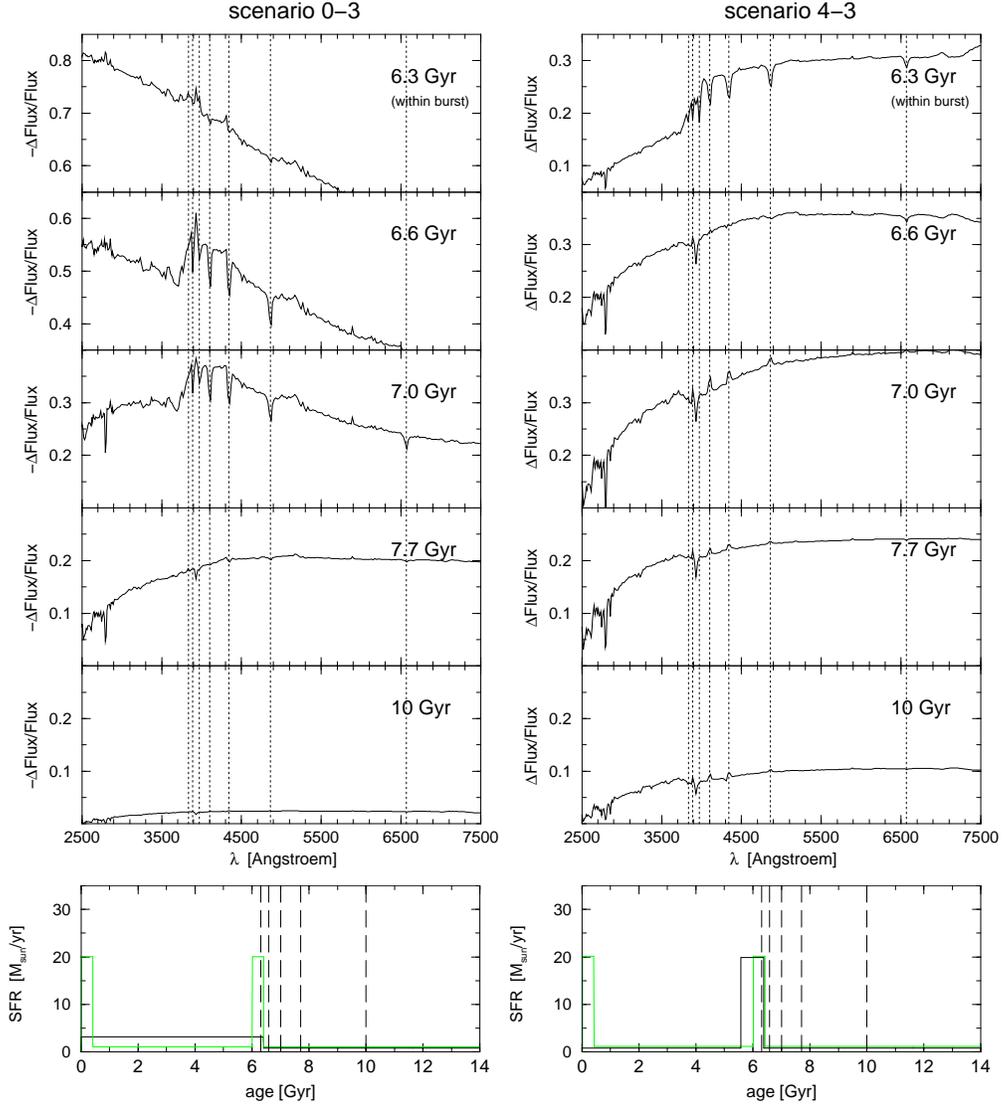}
   \caption{{\itshape Top:\/} Relative differences of the spectra of the scenarios 0 and 3 (left) and of
   the scenarios 4 and 3 (right) within the range of the Balmer lines (with markings of the Balmer lines
   $H_{\alpha} - H_{\eta}$) for the five model ages indicated in the plot.
   {\itshape Bottom:\/} SFH of the respective scenarios; the ages 6.3, 6.6, 7.0, 7.7 and 10 Gyr are indicated by
   dashed lines.}
   \label{abb.spectra}
   \end{center}
\end{figure}

As for colors and indices, scenarios are indistinguishable 4 Gyrs after the latest burst in terms of their full
spectral energy distributions (SEDs) from U through K, as shown in Lilly \& Fritze -- v. Alvensleben (2005b);
however, in contrast to broad-band colors and Lick indices, a clearer distinction between scenarios with bursts of
different durations than between bursts of different strength can be observed in the spectral continua.

Unfortunately, the model spectra we use have a resolution too low to analyse spectral features in detail.
Nevertheless, to get an impression of how different scenarios of SF are reflected in the resulting spectra, in
Figure \ref{abb.spectra} we plot relative difference spectra for two pairs of scenarios for five evolutionary
stages (6.3, 6.6, 7.0, 7.7, and 10 Gyr). In the left panels, a scenario with a weak, uniform SFR (scenario 0) is
confronted with burst scenario 3; in the right panels, this burst scenario 3 is confronted with a scenario
featuring a burst of equal strength but twice the duration (scenario 4).
We only plot a small section of the spectrum from 2500 to 7500 \AA; for better orientation, Balmer lines $H_\alpha$
to $H_\eta$ are marked by vertical lines.

As expected, and as already shown by using Lick indices, the Balmer lines are clearly visible only during a
lookback time of about 1 Gyr after the end of the most recent burst (i.e., they disappear at a galaxy age of 7.7
Gyr); this was already explained by the short lifetimes of stars mainly responsible for these spectral lines.
Other absorption lines easily visible in the plots are the MgII line (at $\lambda$ = 2798.00 \AA) and CaII K (at
$\lambda$ = 3933.44 \AA, between H$_\epsilon$ and H$_\zeta$).
At fixed metallicity, the MgII line is most pronounced in the spectra of early F-type stars with a lifetime of
approx. 2 Gyrs. The CaII K line, on the other hand, is strongest in the spectra of late F-type and early G-type
stars; these stars have lifetimes of 2-6 Gyrs.
These lifetimes explain the relatively complicated behaviour of the lines in the relative difference spectra.

For example, at a galaxy age of 6.6 Gyr (300 Myrs after the end of the last burst), CaII K is \emph{less} deep in
burst scenario 3 than in non-burst scenario 0 (Fig. \ref{abb.spectra}, left panels); about 1 Gyr later, at a galaxy
age of 7.7 Gyr, the line is slightly stronger in the burst scenario:
Shortly after the end of the epoch of enhanced SFR more G-type stars have been accumulated in the non-burst
scenario 0; stars of this type originating from the early burst of scenario 3 are not alive any more at this time.
1 Gyr later, late F- and early G-type stars originating from the most recent burst of scenario 3 outweight the
resp. stars accumulated in the non-burst scenario.\\

Due to the very different lifetimes of stars of various types which are responsible for different spectral lines,
we expect that highly resolved spectra can reveal much more precise information about the recent SFH of galaxies
than colors or the classical set of Lick indices.
A new set of spectral indices with much more narrow passband definitions could provide such a tool. It requires
large telescopes both for the calibrations on stellar spectra and for galaxy observations which will, however,
still be limited to reasonable small distances.\\
%

\section{Summary}

In this study, we wanted to understand to what precision star formation histories can be determined using
integrated colors or indices only.
The main questions were: 
What can be revealed by integrated galaxy data about the SFH \emph{before} the most recent burst or epoch of
important SF?
How long is the lookback time to discriminate between different kinds of simplified SFHs, i.e. epochs of constant
low and high SFRs?\\

Colors mainly depend on the amount of blue luminous high-mass stars that formed within the last 0.5 - 1 Gyr, and
the relation of this amount to that of red low-mass stars formed in earlier epochs of star formation.
Therefore, it is possible to distinguish between different burst strengths (rather than burst durations) within a
lookback time of only about 1, at the utmost 4 Gyrs.

Our simulations show that scenarios featuring bursts of equal strengths but different durations, or different SFHs
before the most recent burst are practically indistinguishable directly after the end of the most recent burst.
Scenarios featuring bursts of different strengths (and different durations), on the other side, allow for a
lookback time of about 1 Gyr. After that time, the scenarios are nearly indistinguishable in praxi
($\Delta_{color}$ $\le$ 0.1mag); about 4 Gyrs after the end of the last burst, all scenarios have identical colors
within realistic observational uncertainties.\\

In terms of metall-sensitive, age-insensitive Lick indices, scenarios with bursts of equal strength but
different durations or former SFH are practically indistinguishable already within the burst; scenarios featuring
bursts of different strengths can be discriminated for no longer than 2 Gyrs after the end of the burst.
Age-sensitive indices allow for longer lookback times with a much larger difference in index strength between
different scenarios; 4 Gyrs after the end of the latest burst, however, all scenarios are, in practice,
indistinguishable in all indices.
Remarkably, this is the same maximal lookback time as for broad-band colors.\\

As for colors and indices, in terms of SEDs from U through K it is not possible to discriminate between different
scenarios 4 Gyrs after the most recent burst.
However, relative difference spectra reveal that individual spectral lines can trace the formation of stars much
better:

The strengths of individual absorption lines in integrated spectra depend on the relative amount of stars in the
stellar population which are mainly responsible for the respective line, i.e. in whose spectra the line is strong
compared to the spectra of other types of stars. Hence, the lookback time allowed by individual lines mainly
depends on the very different lifetimes of these stars.
High resolution spectra or newly defined sets of spectral indices with much more narrow passband definitions may
therefore allow to reveal much more precise information about the SFH of galaxies than our present study can
show.\\
%

\section{Outlook}


The study presented here is part of our preparatory work within a larger collaborative project,
which aims at confronting different methods to derive SFHs directly.

Test object for this ongoing project is an LMC bar field, for which both an integrated spectrum (obtained with
the 3.6m ESO telescope, LaSilla) and data on its resolved stellar population (obtained with the \emph{Hubble
Space Telescope}) are available. That way, the results of the different groups analysing the integrated
spectrum cannot only be compared with each other but also be compared with the SFH obtained by an analysis of
the CMD of the same field.

A short description of the project, as well as details about its observational parts, can be found in Alloin
et al. (2002); an analysis of the CMD for this field is presented by Smecker-Hane et al. (2002).\\

In an accompanying paper (Lilly \& Fritze-v. Alvensleben 2005a, these proceedings) we apply the methods presented
in this paper to the integrated light spectrum of this field and investigate to what extent its SFH can be
constrained on the basis of data available.\\
%

\acknowledgements             
TL gratefully acknowledges partial travel support from the organizers; 
his work is partially funded by DFG grant Fr 916/11-1-2-3.\\


\end{document}